\documentclass[twocolumn,showpacs,aps]{revtex4}
\usepackage{graphicx}
\usepackage{amsmath,amssymb}

\begin{document}
\title{%
Realization of holonomic single-qubit operations
}
\author{K. Toyoda, K. Uchida, A. Noguchi, S. Haze, and S. Urabe}
\affiliation{%
Graduate School of Engineering Science, Osaka University,
1-3 Machikaneyama, Toyonaka, Osaka, Japan
}
\date{\today}

\begin{abstract}
Universal single-qubit operations based on purely geometric phase
factors in adiabatic processes 
are demonstrated by utilizing a four-level system in a trapped
single $^{40}$Ca$^+$ ion connected by three oscillating fields.
Robustness against parameter variations is studied.  The
scheme demonstrated here can be employed as a building block for
large-scale holonomic quantum computations, which may be useful for
large qubit systems with statistical variations in system parameters.
\end{abstract}


\maketitle

\section{Introduction}

Experimental studies of quantum information processing (QIP) has
progressed much in recent years.  In studies of QIP using trapped ions,
realization of small scale computation and entanglement generation with
relatively high fidelity has been reported 
\cite{Home2009}, which reveals that there is
no fundamental obstacle to scaling a large number of ions.  The
upcoming challenges are large-scale operations and high-fidelity gate
operations toward fault-tolerant quantum computation.

As a way to realize high-fidelity gate operations, quantum gates and
quantum computation using geometric phase factors have recently been
studied.  This originates from Holonomic Quantum Computation (HQC)
proposed by Zanardi {\it et al.}\cite{Zanardi1999} \ In HQC, degenerate
multiple quantum states are utilized and unitary operations are
performed by varying the system Hamiltonian along a closed path in the
parameter space.  The final state in the HQC is dependent only on the
global property of the closed path; therefore, HQC is considered to be
robust against certain types of errors.  Even when diabatic evolutions
of the system or nondegenerate quantum states are used, similar
advantages can be expected as long as the unitary operations performed
are determined by geometric phase factors [Geometric Quantum Computation
(GQC)\cite{GQC}].

There have been a number of experiments in different systems related to
GQC \cite{Jones2000,Leibfried2003,Tian2004,Imai2008,Benhelm2008}.  It is
known that universal quantum computation can be realized with a
combination of single-qubit and two-qubit operations
\cite{Barenco1995b}.  As demonstrations of two qubit operations in
trapped-ion systems, ``Geometric-phase gate'' by Leibried {\it et al.}\
\cite{Leibfried2003} and the M\o{}lmer-S\o{}rensen gate
\cite{Molmer1999,Sackett2000,Benhelm2008} have been realized.  A gate
fidelity of 99.3\% has been realized using such a scheme
\cite{Benhelm2008}.

Single qubit operations in trapped-ion systems have been performed to
date using a dynamical method with variable pulse lengths and phases.
By replacing such a method with those that use geometric phases, gate
operations that are robust against variation of parameters such as pulse
intensity and lengths can be expected.

In this work, we report the realization of purely geometric single qubit
operations using a four-level system in a single $^{40}$Ca${^+}$ ion.  
Three transitions in the four-level system are excited either by
three optical fields or by two optical fields and one RF magnetic field.
Rotation operations by arbitrary angles along two different axes, $X$
and $Z$ in the Bloch sphere for the qubit, are demonstrated by utilizing stimulated Raman adiabatic passage (STIRAP) in this four-level system.
Robustness against parameter variations is also demonstrated.  

This work is based on proposals using a {\it tripod} system comprising
of one upper state and three lower states connected by three oscillating
fields \cite{Unanyan1998,Duan2001,Kis2002,Moller2007}.  There are two
dark states in the system, and those dark states are
adiabatically manipulated with the intensities and phases of the
oscillating fields to perform single-qubit operations.

There was also a proposal and demonstrations of single-qubit operations
using geometric phase factors using a two-level system and square pulses
\cite{Tian2004,Imai2008}.  However, to the best of our knowledge no
adiabatic (or holonomic) demonstration of single-qubit operations
using dark states has been reported so far.

\section{Principles for holonomic single-qubit operations}

\begin{figure}[htbp]
\includegraphics[width=8cm]{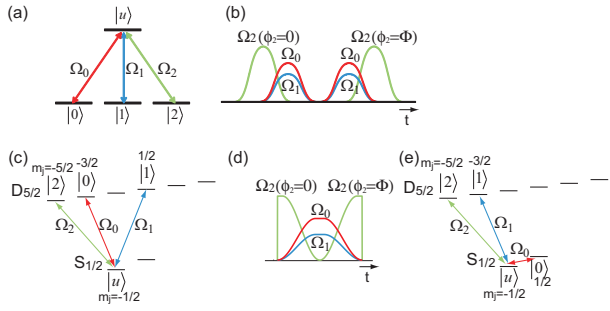} 
\caption{
(Color online)
(a) Level scheme of a four-level system used to implement 
holonomic single-qubit operations.
(b) Pulse sequence for holonomic single-qubit operations.
(c) Level scheme used for $^{40}$Ca$^+$ in the experiment.
(d) Experimentally used pulse sequence.  
(e) Level scheme that implements another four-level system
containing one RF and two optical transitions,
with qubit levels ${\left|0\right\rangle,\left|1\right\rangle}$ 
separated by an optical frequency.
} \label{fig:fig1}
\end{figure}

Here the formalization of the scheme is summarized based on Kis and
Renzoni \cite{Kis2002} for later reference.
A four-level system comprising of
$\{\left|0\right\rangle,\left|1\right\rangle,\left|2\right\rangle,\left|u\right\rangle\}$
is considered
[see Fig.\ \ref{fig:fig1}(a)].
$\left|0\right\rangle$ and $\left|1\right\rangle$ span a qubit
manifold, while $\left|u\right\rangle$ represents an upper state 
and $\left|2\right\rangle$ represents an auxiliary state used in STIRAP.
Three near resonant oscillating fields are applied to this
system between $\left|u\right\rangle$ and $\left|k\right\rangle
(k=0,1,2)$, and the Hamiltonian in the interaction picture with the
rotating wave approximation is given as
\begin{equation}
H(t)=-\hbar\delta \left|u\right\rangle\left\langle{}u\right|
+\frac{\hbar}{2}\sum^{2}_{k=0}
\left[\Omega_k(t)\left|u\right\rangle\left\langle{}k\right|
+{\rm H.c.}\right],
\end{equation}
where $\delta$ is the detuning of the lasers, which is assumed to be
common to all the three transitions.  The Rabi frequencies for the
transitions between $\left|u\right\rangle$ and the qubit states are
$\Omega_0(t)\equiv\Omega(t)\cos\theta_1$ (assumed to be real) and
$\Omega_1(t)\equiv\Omega(t)e^{i\phi_1}\sin\theta_1$, where a real quantity
$\Omega(t)$ is the common envelope function.  
$\theta_1=\tan^{-1}|\Omega_1(t)/\Omega_0(t)|$ and
$\phi_1=\arg\Omega_1(t)$ 
are assumed to be independent of time.
The Rabi frequency between $\left|u\right\rangle$ and
$\left|2\right\rangle$ is described as
$\Omega_2(t)\equiv|\Omega_2(t)|e^{i\phi_2(t)}$ with
$\phi_2(t)=\arg\Omega_2(t)$.

The following states
in the qubit manifold are
useful for describing the Hamiltonian and explaining the gate procedure:
 $\left|C_1\right\rangle=\cos\theta_1\left|0\right\rangle
  +e^{-i\phi_1}\sin\theta_1\left|1\right\rangle$ and 
 $\left|D_1\right\rangle=-\sin\theta_1\left|0\right\rangle
  +e^{-i\phi_1}\cos\theta_1\left|1\right\rangle$,
and similarly, the following states in the total lower-state
manifold spanned by
$\{\left|0\right\rangle,\left|1\right\rangle,\left|2\right\rangle\}$:
 $\left|C_2\right\rangle=
  \cos\theta_2(t)\left|C_1\right\rangle+e^{-i\phi_2(t)}\sin\theta_2(t)\left|2\right\rangle$
  and
 $\left|D_2\right\rangle=
  -\sin\theta_2(t)\left|C_1\right\rangle+e^{-i\phi_2(t)}\cos\theta_2(t)\left|2\right\rangle$,
where $\theta_2(t)=\tan^{-1}|\Omega_2(t)/\Omega(t)|$.
Based on these, the Hamiltonian can be simplified as follows:
\begin{equation}
 H(t)=-\hbar\delta \left|u\right\rangle\left\langle{}u\right|
+\frac{\hbar}{2}
\Omega_{\mathrm {total}}(t)
\left(
\left|u\right\rangle\left\langle{}C_2\right|+{\rm H.c.}
\right),
\label{eq:H2}
\end{equation}
where $\Omega_{\mathrm {total}}(t)\equiv\left[\Omega^2(t)+|\Omega_2(t)|^2\right]^{1/2}$.
When $\delta=0$, this Hamiltonian has four eigenvectors
$\{\left|D_1\right\rangle,\left|D_2\right\rangle,\left|B_+\right\rangle,\left|B_-\right\rangle\}$
with eigenvalues of $\{0,0,\hbar\Omega_{\mathrm
{total}}(t)/2,-\hbar\Omega_{\mathrm {total}}(t)/2\}$, respectively,
where
$\left|B_\pm\right\rangle=(\left|C_2\right\rangle\pm\left|u\right\rangle)/\sqrt{2}$.

We proof the ability to perform arbitrary single-qubit operations
by first starting from an arbitrary
initial state in the qubit manifold,
$\left|\psi_0\right\rangle\equiv\alpha\left|0\right\rangle+\beta\left|1\right\rangle$,
which can be rewritten using $\left|C_1\right\rangle$ and
$\left|D_1\right\rangle$ as $\left|\psi_0\right\rangle=
\left\langle{}C_1|\psi_0\right\rangle\left|C_1\right\rangle+
\left\langle{}D_1|\psi_0\right\rangle\left|D_1\right\rangle$.  This initial
state will
be transferred in the first STIRAP sequence 
that makes use of $\left|D_2\right\rangle$ 
[the left half in Fig.\
\ref{fig:fig1}(b), with $\theta_2(t):\pi/2\rightarrow0$ and $\phi_2=\arg\Omega_2=0$]
to the intermediate state
$\left|\psi_1\right\rangle=
-\left\langle{}C_1|\psi_0\right\rangle\left|2\right\rangle+
\left\langle{}D_1|\psi_0\right\rangle\left|D_1\right\rangle$.  This
intermediate state will then
be transferred using the second STIRAP sequence [the right half in
Fig.\ \ref{fig:fig1}(b), with $\theta_2(t):0\rightarrow\pi/2$ and
$\phi_2=\arg\Omega_2=\Phi$] to the final state $\left|\psi_2\right\rangle=
e^{i\Phi}\left\langle{}C_1|\psi_0\right\rangle\left|C_1\right\rangle+
\left\langle{}D_1|\psi_0\right\rangle\left|D_1\right\rangle$,
which can be rewritten
using the identity operator $\hat{I}$ and the Pauli operators $\hat{\rm\bf \sigma}=(\hat\sigma_x,\hat\sigma_y,\hat\sigma_z)$
on the computational manifold $\{\left|0\right\rangle,\left|1\right\rangle\}$
as 
$\left|\psi_2\right\rangle=e^{i\Phi/2}[\cos(\Phi/2)\hat{I}+i{\rm\bf
n}\cdot\hat{\rm\bf \sigma}\sin(\Phi/2)]\left|\psi_0\right\rangle$ 
with 
${\rm\bf n}=(\sin2\theta_1\cos\phi_1,-\sin2\theta_1\sin\phi_1,\cos2\theta_1)$.
This operation corresponds to a rotation by angle $-\Phi$ around
${\rm\bf n}$, which can be taken arbitrarily by selecting the values for
$\theta_1$ and $\phi_1$, hence arbitrary single-qubit operations can be
performed.

This scheme is based on adiabatic population transfer using
$\left|D_2\right\rangle$, and diabatic transitions 
from this state to 
$\left|B_\pm\right\rangle$ are among the possible causes of infidelity
in gate operations.
The probability for such diabatic transitions
is calculated to be in the order of
$\max\left[2\dot{\theta_2(t)}^2/\Omega_\mathrm{total}^2\right]$%
\cite{Messiah1961},
which should be set to be much smaller than 1 to maintain
high fidelity.

\section{Experimental setup and procedures}

The experimental setup has been previously described \cite{Toyoda2011} and 
only a brief description is given here. A single $^{40}$Ca$^+$ is trapped in vacuum
($6\times10^{-9}$ Pa) using a linear Paul trap.  
The trap used here is a conventional linear trap with an operating
frequency of 23 MHz and secular frequencies of
$(\omega_x,\omega_y,\omega_z)/2\pi=(2.4,2.2,0.69)$ MHz.
A bias magnetic field of
$2.9\times10^{-4}$ T is applied to define a quantization axis, which
results in a Zeeman splitting of $\sim$4.9 MHz between $D_{5/2}$
sublevels.  
In experiments that employ three optical fields, 
a titanium sapphire laser at 729 nm stabilized to a
high-finesse low-thermal-expansion cavity is used for the excitation of
ions between $S_{1/2}$ and $D_{5/2}$.  The amplitudes and frequencies of
the three optical fields at 729 nm are changed by varying the three RF
fields that are combined and fed to an acousto-optic modulator.
The RF fields are generated by three
direct-digital synthesis (DDS) boards that are controlled by a
field-programmable gate array.  Polarization of the optical
fields is adjusted so that there are polarizations both parallel and
perpendicular to the bias magnetic field; therefore, the transition to be excited
is selected by changing the frequencies of the DDS boards.
In experiments that employ two optical fields and one RF magnetic field,
an RF coil in the vicinity of the trap is used to generate the RF field.
The details of the RF excitation procedure are similar to that 
described in Ref.\ \cite{Ohno2011}.

We have chosen the $S_{1/2}$--$D_{5/2}$ electric-quadrupole transition
of $^{40}$Ca$^+$ for realizing the geometric phase gate.
The encoding of the tripod system to the sublevels in $S_{1/2}$ and
$D_{5/2}$
is as follows: 
$S_{1/2}(m_j=-1/2)$ as
the ``upper'' state $\left|u\right\rangle$ and three Zeeman sublevels in
$D_{5/2}$ ($m_j=-3/2, 1/2, -5/2$) as the lower states
$\left|0\right\rangle$, $\left|1\right\rangle$ and
$\left|2\right\rangle$, respectively (see Fig.\ \ref{fig:fig1}(c)).  The
reason that a level in $S_{1/2}$ is adopted instead of one in $P_{3/2}$ as
$\left|u\right\rangle$ is to avoid the effects of spontaneous emissions when
all the fields are resonant.  However, unwanted
couplings between $S_{1/2}(m_j=1/2)$ and $\left|0\right\rangle$,
$\left|1\right\rangle$ and $\left|2\right\rangle$ must still be
considered.
Since these
couplings are off resonance, varying $\theta_2(t)$ leads to
time-dependent AC-Stark shifts that disturb the null eigenenergies of
the qubit states.  This effect was numerically evaluated with realistic
parameters and the phase accumulated during the gate
operation was confirmed to be smaller than 0.5 rad under typical experimental conditions in
the present work.  

We essentially do not compensate such AC-Stark shifts for the present
work. In the experiments in \S\ref{sec:xzrotations} and of rotations by
angle $\pi$ in \S\ref{sec:fidelity}, we adjust one of the detunings
(one for $\left|u\right\rangle$--$\left|0\right\rangle$) so that fringe
shift resulting from AC-Stark shifts are apparently canceled.
%
%
This cancelation depends on the details of the STIRAP pulses, including
such conditions as peak Rabi frequencies, pulse shapes or total time
(it may not depend on such conditions as input states or rotation
angles since AC Stark shifts do not depend on these).  We also evaluate
gate performances without cancellation of this type by examining
rotations by angle $\pi/2$ with relatively low Rabi frequencies in
\S\ref{sec:fidelity} and \S\ref{sec:discussions}.

%
%

The procedure for gate operations is as follows.  Firstly, the ion is
cooled to near the motional ground state with Doppler cooling by 397 and
866 nm and with sideband cooling of the axial motion by lasers at 729
and 854 nm.
The average motional quantum number along axial direction
after sideband cooling is $\bar{n}_z=0.06$.
Optical pumping using a 397 nm $\sigma^-$ transition is then performed
to initialize the ion in $S_{1/2}(m_j=-1/2)$ ($\left|u\right\rangle$).
The ion is transferred from $\left|u\right\rangle$ to the computational
manifold spanned by $\{\left|0\right\rangle,\left|1\right\rangle\}$ using
square $\pi$ pulses, and then a gate STIRAP pulse sequence is applied between
$\left|0\right\rangle$ and $\left|1\right\rangle$.  The pulse shape of
the gate consists of partially overlapping sinusoidal
curves and constant values, as in Fig. \ref{fig:fig1}(d).  
A step variation of $\phi_2(t)$ by $\Phi$, which brings a
geometric phase factor, is given at the middle of the
gate sequence where $|\Omega_2(t)|=0$.  After the gate
sequence, a square $\pi$ pulse is applied to map the state of
$\left|0\right\rangle$ to $\left|u\right\rangle$.  
An additional pulse may also be applied, depending on the element of the density
matrix to be observed. State discrimination is performed by detecting
fluorescence from the ion with a photomultiplier tube during a period of
7 ms when 397 and 866 nm fields are applied\cite{Dehmelt1975}.

The time dependence of the optical pulses is chosen as follows 
($t_1=0$, $t_1=(1-\alpha)T$, $t_2=T$ and $t_3=(2-\alpha)T$ 
where $0\le{}\alpha\le{}1$):
\begin{equation}
\Omega^2(t)=
\begin{cases}
0 &(t<t_0)\\
\Omega^2_{\rm max}(1-\cos^2{\pi{}t/2T})&(t_0\le{}t<t_2)\\
\Omega^2_{\rm max}&(t_2\le{}t<t_3)\\
0&(t\ge{}t_3),
\end{cases}
\end{equation}
\begin{equation}
|\Omega_2(t)|^2=
\begin{cases}
0&(t<t_0)\\
\Omega^2_{\rm 2,max}   &(t_0\le{}t<t_1)\\
\Omega^2_{\rm 2,max}\cos^2\pi{}(t-t_1)/2T    & (t_1\le{}t<t_3)\\
0&(t\ge{}t_3),
\end{cases}
\end{equation}
where $\Omega^2_{\rm max}$ and $\Omega^2_{\rm 2,max}$ are the maximum
values for $\Omega^2(t)$ and $|\Omega_2(t)|^2$, respectively.
In the experimental results given in this article,
the values of $\alpha$ were empirically chosen to in between 0.55 and 0.85
so that effective population transfer was attained in experiments
and/or numerical simulations.
It is noted that the values of $\alpha$ used in this article
are not thoroughly optimized ones
but those that give relatively good performances among tested.

\section{Results for $X$- and $Z$-rotations}
\label{sec:xzrotations}
We show that arbitrary one-qubit operations can be realized with the
geometric method introduced here. This can be accomplished by showing
$X$- and $Z$-rotations independently, because arbitrary operations
can be decomposed into sequences of such rotations.

The procedure for $X$-rotations is as follows.  The ion is first prepared
in $\left|u\right\rangle$ by sideband cooling and optical pumping, then
initialized to $\left|0\right\rangle$, and a STIRAP sequence is applied
to perform a geometric gate.  After this, the populations of the final
states are analyzed by mapping the qubit states to the optical
transition using a rectangular $\pi$ pulse on
$\left|i\right\rangle\leftrightarrow\left|u\right\rangle$ ($i=0,1,2$).
The populations in $\left|2\right\rangle$ and $\left|u\right\rangle$ are
also measured to verify that the gate operation is closed to the
manifold spanned by the qubit states
$\{\left|0\right\rangle,\left|1\right\rangle\}$.  $\Omega_0$ and
$\Omega_1$ are adjusted to be equivalent, and the peak values
for $(\Omega_0,\Omega_1,\Omega_2)/2\pi$ are $\sim(125,125,170)$ kHz.
The total time for the STIRAP sequence is set to be $\sim$120 $\mu$s 
by taking appropriate account of the adiabaticity.
$\alpha=0.85$ is used for the results
given in this section.

\begin{figure}[htbp]
\includegraphics[width=8cm]{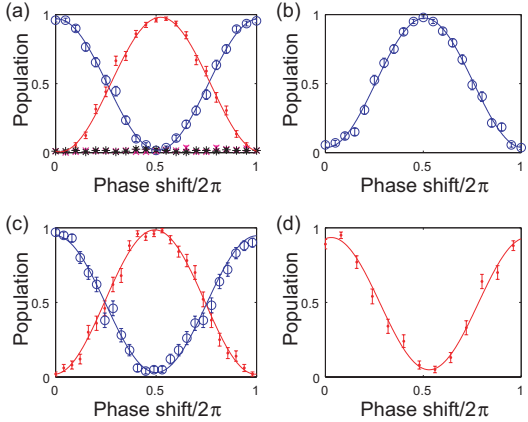}
\caption{%
(Color online)
Results of single-qubit rotations ($X$- and $Z$-rotations).
Variations of population are plotted against the phase shift in the middle of the
gate pulse.
The number of experiments per data points is 200 for experiments using three optical transitions [(a) and (b)] and
100 for those using one RF and two optical transitions [(c) and (d)].
The error bars represent errors in projection measurements,
which are derived as standard deviations in binomial distributions with numbers of samples as just given.
(a) Results for $X$-rotations using three optical transitions.
Blue hollow circles, red filled circles, magenta crosses and black
 asterisks represent the populations in
$\left|0\right\rangle$, $\left|1\right\rangle$, $\left|2\right\rangle$ and
 $\left|u\right\rangle$, respectively.  
(b) Results for $Z$-rotations using three optical transitions.
The population in $\left|0\right\rangle$ is plotted.
(c) Results for $X$-rotations using one RF and two optical transitions. 
Red filled circles (blue hollow circles) represent the populations in
$\left|1\right\rangle$ and $\left|2\right\rangle$ when $\left|0\right\rangle$ 
($\left|1\right\rangle$) is initially prepared.
(d) Results for $Z$-rotations using one RF and two optical transitions.
The population in $\left|1\right\rangle$ and $\left|2\right\rangle$ is plotted.}
\label{fig:xzrotations}
\end{figure}

Figure\ \ref{fig:xzrotations}(a) shows the results for $X$-rotations.  
A sinusoidal oscillation of population between $\left|0\right\rangle$ and
$\left|1\right\rangle$ is observed.
The sum of populations in the other
states $\left|2\right\rangle$ and $\left|u\right\rangle$ is below 0.05 over the
entire region, with which we confirm that the
rotation operations are almost limited to the computational manifold spanned by
$\{\left|0\right\rangle,\left|1\right\rangle\}$.
The contrasts of the populations of $\left|0\right\rangle$ and
$\left|1\right\rangle$ are obtained by fitting to 
$0.951\pm0.009$ and $0.975\pm0.008$, respectively.

The procedure for observation of $Z$-rotations is similar to Ramsey
interferometry. A superposition of the qubit states
$\left|0\right\rangle$ and $\left|1\right\rangle$ is first prepared by
application of 
a $\pi/2$ pulse on $\left|u\right\rangle$--$\left|1\right\rangle$
and a $\pi$ pulse on $\left|u\right\rangle$--$\left|0\right\rangle$.
A $Z$-rotation is then performed with STIRAP, which is followed by a $\pi$ pulse on
$\left|u\right\rangle$--$\left|0\right\rangle$ and a $\pi/2$ pulse on
$\left|u\right\rangle$--$\left|1\right\rangle$.  The last pulse causes an
interference between $\left|u\right\rangle$ and $\left|1\right\rangle$,
which is detected by a projection measurement using fluorescence.  Here
the peak values for $(\Omega_0,\Omega_1,\Omega_2)/2\pi$ are set to be
$\sim$$(200,0,170)$ kHz and the total time for the STIRAP sequence is
$\sim$120 $\mu$s.
Figure \ref{fig:xzrotations}(b) represents the
results for $Z$-rotations, which shows a population oscillation against
$\Phi$ with an almost unit contrast.
From a least-squares fit, the contrast is obtained to be $0.937\pm0.010$.

\section{Evaluation of fidelity}
\label{sec:fidelity}
In order to investigate the action of gate operations more
quantitatively, we also performed estimation of gate fidelities by using
quantum state tomography.  We performed fidelity estimation in the
following three cases:
(a) X-rotation by angle $\pi$,
(b) Z-rotation by angle $\pi$, and
(c) Hadamard gate [rotation around ${\bf
n}=\left(1/\sqrt{2},0,-1/\sqrt{2}\right)$ in the Bloch sphere by angle
$\pi$].
As explained before, general operations of the STIRAP gate can be
described as the following unitary operator:
\begin{eqnarray}
\lefteqn{\hat{U}_{\rm STIRAP}(\theta_1,\phi_1,\Phi)}\nonumber\\
&\equiv{}&e^{i\Phi/2}
\left[
\cos\frac{\Phi}{2}\hat{I}+i{\rm\bf n}(\theta_1,\phi_1)\cdot\hat{\rm\bf
\sigma}\sin\frac{\Phi}{2}
\right]
\end{eqnarray}
where
${\rm\bf n}(\theta_1,\phi_1)$$\equiv$$(\sin2\theta_1\cos\phi_1,$$-\sin2\theta_1\sin\phi_1,$$\cos2\theta_1)$. 
Using this general expression, the unitary operators for the three cases given above are
respectively written as follows:
\begin{eqnarray}
 \hat{U}_X&\equiv&\hat{U}_{\rm STIRAP}(\pi/4,0,\pi),\\
 \hat{U}_Z&\equiv&\hat{U}_{\rm STIRAP}(0,0,\pi),\\
 \hat{U}_H&\equiv&\hat{U}_{\rm STIRAP}(3\pi/8,0,\pi).
\end{eqnarray}

The initial states in the three cases are respectively prepared as
follows:
(a)$\left|\psi_{0X}\right\rangle\equiv\left|0\right\rangle$,
(b)$\left|\psi_{0Z}\right\rangle\equiv(-i\left|0\right\rangle+\left|1\right\rangle)/\sqrt{2}$,
and
(c)$\left|\psi_{0H}\right\rangle\equiv\left|0\right\rangle$.
In order to take into account imperfect preparation of these initial states, we
performed measurement of these states using the technique of quantum
state tomography.  In the encoding adopted here, both the two states
$\left|0\right\rangle$ and $\left|1\right\rangle$ in the computational
subspace are in the same electronic state $D_{5/2}$.  In order to
discriminate these two states, we applied a mapping pulse between
$\left|0\right\rangle$ and $\left|u\right\rangle$ before performing
fluorescence detection.  The population out of the computational
manifold (population in $\left|1\right\rangle$ and
$\left|u\right\rangle$) before application of the mapping pulse, which
is typically below 0.05 as described before, is ignored here for
simplicity.  The mapping pulse itself produces a geometric phase
relative to the other states ($\left|1\right\rangle$ and
$\left|2\right\rangle$), which should be taken into account properly in the
fidelity analysis.  The mapping operation can be described as follows:
\begin{equation}
{\hat{R}_{{\rm SWAP},0u}}
\equiv
-i\left|0\right\rangle \left\langle{}u\right|
-i\left|u\right\rangle \left\langle{}0\right|
+\left|1\right\rangle \left\langle{}1\right|
+\left|2\right\rangle \left\langle{}2\right|.
\end{equation}
This corresponds to a
rotation by angle $\pi$ around the $x$ axis in the Bloch sphere of the
two-level system $\{\left|0\right\rangle, \left|u\right\rangle\}$.

The density matrix after application of the mapping pulse were
reconstructed (using linear reconstruction) by fluorescence detection and
an optional $\pi/2$ pulse prior to that.  The populations along three
orthogonal axes in the Bloch sphere, $P_{x1i}$,$P_{y1i}$ and $P_{z1i}$,
are measured, where $P_{x1i}$ ($P_{y1i}$) is the population in $D_{5/2}$
after the mapping pulse and a $\pi/2$ pulse with the phase $-3\pi/2$
($0$) on $\left|u\right\rangle$--$\left|1\right\rangle$, and $P_{z1i}$
is the population in $D_{5/2}$ state immediately after the mapping
pulse.  Using the Bloch vector ${\bf
r}$$=(r_x,r_y,r_z)$$=(1-2P_{x1i},1-2P_{y1i},1-2P_{z1i})$ the density
operator
after the mapping pulse
is expressed as
\begin{eqnarray}
\lefteqn{\hat{\rho}_{i,obs}}\nonumber\\
&\equiv&\frac{1}{2}
\hat{I}
+\frac{r_x}{2}(  \left|u\right\rangle\left\langle{}1\right|+ \left|1\right\rangle\left\langle{}u\right|)
+\frac{r_y}{2}(-i\left|u\right\rangle\left\langle{}1\right|+i\left|1\right\rangle\left\langle{}u\right|)\nonumber\\
&&+\frac{r_z}{2}(  \left|u\right\rangle\left\langle{}u\right|- \left|1\right\rangle\left\langle{}1\right|),
\end{eqnarray}
and by using this the initial density operator is described as follows:
\begin{equation}
{\hat{\rho}_i}
\equiv
\hat{R}_{{\rm SWAP},0u}^\dagger
\hat{\rho}_{i,obs}
\hat{R}_{{\rm SWAP},0u}.
\end{equation}
In a similar way, the final density operator just after STIRAP gate
operations, $\hat\rho_f$, is obtained in terms of populations 
$P_{x1f}$,$P_{y1f}$ and $P_{z1f}$ (defined in the same way as above) after a mapping pulse
that follows the gate operation.

The gate fidelities in the three cases can be described
using the density operators and the unitary operators, as follows:
\begin{eqnarray}
 F_X&\equiv&{\rm tr}\left(\hat\rho_f\hat{U}_X\hat\rho_i\hat{U}_X^\dagger\right),\\
 F_Z&\equiv&{\rm tr}\left(\hat\rho_f\hat{U}_Z\hat\rho_i\hat{U}_Z^\dagger\right),\\
 F_H&\equiv&{\rm tr}\left(\hat\rho_f\hat{U}_H\hat\rho_i\hat{U}_H^\dagger\right).
\end{eqnarray}
The fidelities are explicitly written in terms of the initial
 and final populations, as follows:
\begin{eqnarray}
 F_X&=& P_{x1i}+P_{x1f}-2P_{x1i}P_{x1f}\nonumber\\
    & &-P_{y1i}-P_{y1f}+2P_{y1i}P_{y1f}\nonumber\\
    & &+P_{z1i}+P_{z1f}-2P_{z1i}P_{z1f},\\
 F_Z&=& P_{x1i}+P_{x1f}-2P_{x1i}P_{x1f}\nonumber\\
    & &+P_{y1i}+P_{y1f}-2P_{y1i}P_{y1f}\nonumber\\
    & &-P_{z1i}-P_{z1f}+2P_{z1i}P_{z1f},\\
 F_H&=& P_{x1i}+P_{x1f}-2P_{x1i}P_{x1f}\nonumber\\
    & &+P_{y1i}+P_{z1f}+2P_{y1i}P_{z1f}\nonumber\\
    & &+P_{z1i}+P_{y1f}-2P_{z1i}P_{y1f}-1.
\end{eqnarray}
In order to determine the confidence intervals of the fidelities, we
obtained variances of the fidelities considering propagation of
uncertainty based on the above expression, as follows:
\begin{eqnarray}
\lefteqn{V(F_X)}\nonumber\\
&=&V(P_{x1i})+V(P_{x1f})+4P_{x1i}^2V(P_{x1f})+4P_{x1f}^2V(P_{x1i})\nonumber\\
       & &V(P_{y1i})+V(P_{y1f})+4P_{y1i}^2V(P_{y1f)}+4P_{y1f}^2V(P_{y1i)}\nonumber\\
       & &V(P_{z1i})+V(P_{z1f})+4P_{z1i}^2V(P_{z1f})+4P_{z1f}^2V(P_{z1i}),\nonumber\\
\lefteqn{V(F_Z)=V(F_X),}\nonumber\\
\lefteqn{V(F_H)}\nonumber\\
&=&V(P_{x1i})+V(P_{x1f})+4P_{x1i}^2V(P_{x1f})+4P_{x1f}^2V(P_{x1i})\nonumber\\
       & &V(P_{y1i})+V(P_{z1f})+4P_{y1i}^2V(P_{z1f})+4P_{z1f}^2V(P_{y1i})\nonumber\\
       & &V(P_{z1i})+V(P_{y1f})+4P_{z1i}^2V(P_{y1f})+4P_{y1f}^2V(P_{z1i}),\nonumber
\end{eqnarray}
where $V(...)$ represents the variance of each quantity.  The variances
of the populations [such as $V(P_{x1i})$] are simply determined here as
the variances in binomial distributions; {\it e.g.}
$V(P_{x1i})=P_{x1i}(1-P_{x1i})/N$, where $N$ is the number of
experiments per one measurement condition.  The confidence intervals
(68\%) for the fidelities are determined as the square roots of the
variances of the fidelities.

Table \ref{tab1} shows the results for the initial and final population
measurements. The measurements were performed in essentially the same
conditions as used in the previous section.  Using the measured
populations, the fidelities are estimated and shown in Table \ref{tab2}.
The second column in Table \ref{tab2} shows the fidelities reflecting
both initial and final populations.  We should note that these values
are largely affected by imperfect initialization and analysis and
therefore should be rather taken as the lower limits for the fidelities.

For the purpose of reference, we also estimated the fidelities when
ideal initial or final states are assumed (the third and fourth columns
in Table \ref{tab2}, respectively).  The fidelities for ideal initial
states (the third column) represent the goodness of the generated states
without considering initialization errors.  These values that are better
than those in the second column can be considered as the upper limits
for the fidelities (when all operations other than the gates are assumed
to be perfect).  The fidelities for ideal final states (the fourth
column) represent the goodness of the initialization process.  These
values support the observation that the values in the second column is
largely affected by imperfect initialization (and possibly imperfect
analysis).

\begin{table}
\begin{tabular}{lllllll}
\toprule Gate
type&$P_{x1i}$&$P_{y1i}$&$P_{z1i}$&$P_{x1f}$&$P_{y1f}$&$P_{z1f}$\\
\colrule 
X-$\pi$ &0.510&0.549&0.026&0.463&0.555&0.985\\
Z-$\pi$ &0.509&0.045&0.502&0.472&0.973&0.499\\
Hadamard&0.497&0.515&0.025&0.485&0.989&0.488\\ 
\botrule
\end{tabular}
\caption{Results of population measurements for fidelity estimation.
The initial ($\{P_{x1i},P_{y1i},P_{z1i}\}$) and final
($\{P_{x1f},P_{y1f},P_{z1f}\}$) populations in the different three bases
are measured for three different types of gate operations (see the text
for details).  Each result is obtained as the average of 1500
experiments.}  \label{tab1}
\end{table}

\begin{table}
\begin{tabular}{llll}
\toprule
Gate type&Fidelity&Fidelity &Fidelity\\
&&(ideal initial&(ideal final\\
&&states)&states)\\
\colrule
X-$\pi$ &0.965$\pm$0.038&0.985$\pm$0.026&0.974$\pm$0.027\\
Z-$\pi$ &0.931$\pm$0.038&0.973$\pm$0.036&0.955$\pm$0.039\\
Hadamard&0.965$\pm$0.038&0.989$\pm$0.026&0.975$\pm$0.027\\
\botrule
\end{tabular}
\caption{Estimated fidelities. The confidence intervals are for 68\%
 confidence level. The second column shows fidelities calculated from
 the measured initial and final populations.  The third and fourth
 columns show estimated fidelities when ideal initial or final states
 are assumed, respectively, for the purpose of reference (see the text
 for details).  } \label{tab2}
\end{table}

We also performed fidelity analysis for STIRAP gates with
rotation angle of $\pi/2$, which are more
sensitive to coherence between STIRAP and square-pulse operations
and hence to AC Stark shifts.
Table \ref{tab3} shows the results for the initial and final population
measurements for X and Z gates.
Either 
$\left|+x\right\rangle\equiv(\left|0\right\rangle+\left|1\right\rangle)/\sqrt{2}$ or
$\left|+z\right\rangle\equiv\left|0\right\rangle$
as specified are used as the initial state.
It should be noted that this time AC Stark shifts are not compensated at all, and
relatively small values of the Rabi frequencies are used to suppress their
effect.
The peak values for $(\Omega_0,\Omega_1,\Omega_2)/2\pi$ are set to be
$\sim$$(33.5,33.7,47.0)$ kHz for X gates and
$\sim$$(47.3,0,47.2)$ kHz for Z gates.
The total time for the STIRAP sequence is
$\sim$290 $\mu$s and $\alpha=0.75$ is used.
The results of population measurements are shown in Table \ref{tab3}.
Using these populations, the fidelities and their 
confidence intervals are estimated 
in the same way as above and shown in Table \ref{tab4}.

It should be noted that the fidelities in the case of
ideal final states in Table \ref{tab4} 
are higher than what are given in Table \ref{tab2}.
When taking the results in Table \ref{tab3},
we used a noise eater with a sample and hold capability
to reduce fluctuations in the amplitudes of the square pulses.
We speculate that this helped improving the operational fidelity
of the initialization.

\begin{table}
\begin{tabular}{lllllll}
\toprule Gate
type&$P_{x1i}$&$P_{y1i}$&$P_{z1i}$&$P_{x1f}$&$P_{y1f}$&$P_{z1f}$\\
\colrule 
X-$\pi/2$($\left|+x\right\rangle$ prep.) & 0.009 & 0.505 & 0.515 & 0.029 & 0.611 & 0.622\\
X-$\pi/2$($\left|+z\right\rangle$ prep.) & 0.526 & 0.540 & 0.006 & 0.595 & 0.916 & 0.363\\
Z-$\pi/2$($\left|+x\right\rangle$ prep.) & 0.009 & 0.505 & 0.515 & 0.592 & 0.055 & 0.515\\
Z-$\pi/2$($\left|+z\right\rangle$ prep.) & 0.526 & 0.540 & 0.006 & 0.546 & 0.534 & 0.041\\
\botrule
\end{tabular}
\caption{
Results of population measurements for fidelity estimation of
 X and Z gates with $\pi/2$ rotations.
Either 
$\left|+x\right\rangle\equiv(\left|0\right\rangle+\left|1\right\rangle)/\sqrt{2}$ or
$\left|+z\right\rangle\equiv\left|0\right\rangle$
as specified
are used as the initial state.
Each result is obtained as the average of 1500
experiments.
Note that
the same results are used as the initial populations for X and Z gates.
}  \label{tab3}
\end{table}

\begin{table}
\begin{tabular}{llll}
\toprule
Gate type&Fidelity&Fidelity &Fidelity\\
&&(ideal initial&(ideal final\\
&&states)&states)\\
\colrule
X-$\pi/2$($\left|+x\right\rangle$ prep.) & 0.960$\pm$0.039 & 0.971$\pm$0.025 & 0.991$\pm$0.026\\
X-$\pi/2$($\left|+z\right\rangle$ prep.) & 0.905$\pm$0.037 & 0.916$\pm$0.026 & 0.994$\pm$0.026\\
Z-$\pi/2$($\left|+x\right\rangle$ prep.) & 0.936$\pm$0.038 & 0.945$\pm$0.026 & 0.991$\pm$0.026\\
Z-$\pi/2$($\left|+z\right\rangle$ prep.) & 0.952$\pm$0.038 & 0.959$\pm$0.026 & 0.994$\pm$0.026\\
\botrule
\end{tabular}
\caption{Fidelities
for gates with $\pi/2$ rotations
estimated from measured populations in the previous table.  } \label{tab4}
\end{table}

\section{Results for qubit encoding to optically separated levels}
\label{sec:rf}
We have also attempted encoding of the qubit into optically separated
levels.  Use of optically separated levels may be advantageous when
considering a combination with optical two-qubit gate schemes \cite{Molmer1999,Sackett2000,Benhelm2008}
and use of
less magnetic-field-sensitive transitions, such as
$S_{1/2}(m_J=1/2)$--$D_{5/2}(m_J=1/2)$.  This encoding uses an RF
transition between the ground Zeeman sublevels \cite{Ohno2011}.
Figure \ref{fig:fig1}(e) shows the levels used for this encoding.
$\{\left|0\right\rangle,\left|1\right\rangle,\left|2\right\rangle,\left|u\right\rangle\}$
are encoded into $S_{1/2}(m_J=1/2)$, $D_{5/2}(m_J=-3/2)$,
$D_{5/2}(m_J=-5/2)$ and $S_{1/2}(m_J=-1/2)$, respectively.

Fig.\ \ref{fig:xzrotations}(c)
shows the results for $X$-rotations with
qubit encoding to optically separated levels.
Blue hollow circles (red filled circles) 
represent the populations in
$\left|0\right\rangle$
when $\left|0\right\rangle$ 
($\left|1\right\rangle$) is initially prepared.
The contrasts are $0.967\pm0.012$ and $0.916\pm0.014$ for
the preparation of $\left|0\right\rangle$ and $\left|1\right\rangle$, respectively.
Fig.\ \ref{fig:xzrotations}(d) shows the result for
$Z$-rotations, with a contrast of $0.886\pm0.022$.
In these results, the peak values of $(\Omega_0,\Omega_1,\Omega_2)/2\pi$ are set to be
$\sim(130,100,100)$ kHz and $\sim(130,0,100)$ kHz for $X$- and $Z$-rotations, respectively.
The total time for the STIRAP sequence is
approximately 128 $\mu$s.  

In ideal cases of $X$-rotations,
the peak values of $\Omega_0$ and $\Omega_1$ should be equal,
which is not the case in the experiment described above.
For a technical reason concerning the difference in the 
amplitude modulation of RF and optical fields, the 
temporal shapes of $\Omega_0$ and $\Omega_1$ are not proportional
to each other in the present setup.
The peak values given above are determined empirically so that
they give correct $X$-rotations.
This imperfection may be avoided by simply calibrating the amplitude
modulation process
for either the RF or optical field so that
$\Omega_0$ and $\Omega_1$ give exactly the same time dependency.

\section{Demonstration of Robustness}

The feature of the geometric phase gate is that it is expected to be
robust against
variations in the pulse area, namely the Rabi frequency and the pulse
length.  To study this feature, two 
experiments were conducted.  
One is the measurement of $X$-rotations with variation of the peak Rabi
frequency and the illumination period.
The other is the measurement of $Z$-rotations with variation of the
peak Rabi frequency,
while the illumination period is held constant.
The ratios of the peak Rabi frequencies are held constant in each case.
Both experiments are performed by
measuring population oscillations as the rotation angle $\phi_2$ is
varied.

\begin{figure}[htbp]
\includegraphics[width=8cm]{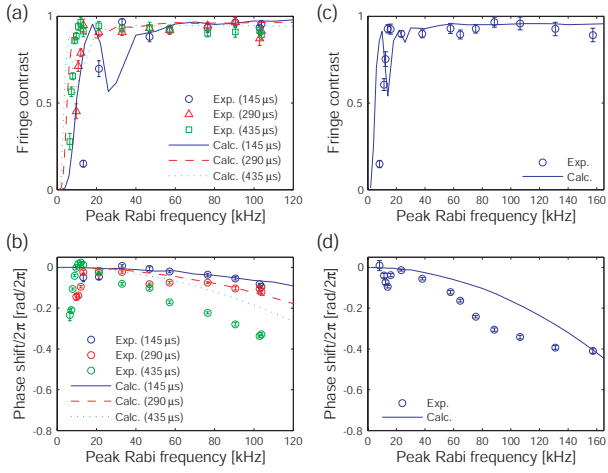} 
\caption{ 
(Color online)
Demonstration of the robustness of single-qubit operations. 
Each point represent the fringe contrast and shift of the population oscillation
obtained by varying the rotation angle of a certain gate operation.
Here the rotation angle is varied from 0 to $2\pi$ in 17 steps, and the
 number
of experiments per angle step is 50.
The errors in the population measurements are estimated in the same way
 as in Fig.\ \ref{fig:xzrotations}, and the fringe contrasts and shifts are
 obtained through weighted fits considering those errors.
The error bars in the figure represents errors in the fitted parameters.
$\alpha=0.55$ is used here.
(a)Fringe contrasts and (b) shifts for $X$-rotations
as the peak Rabi frequencies are varied. Here
the ratios of the peak values are fixed.
The horizontal axis represents the peak value of
$\Omega_2$, and the blue circles, red triangles and green squares represent
the results for pulse durations of 145, 290 and 435 $\mu$s,
 respectively. 
Numerically calculated results for pulse durations of 145, 290 and 435 $\mu$s
are also plotted as
blue solid curves, red dashed curves and green dotted curves, respectively. 
(c) Fringe contrasts and (d) shifts for $Z$-rotations
as the peak Rabi frequencies are varied. 
The ratios of the peak values are fixed.
Numerically calculated results
are also plotted as a blue solid curve.
}
\label{fig:robustness}
\end{figure}

The results for the former experiment is shown in Fig.\
\ref{fig:robustness} (a) and (b)
for three values of the illumination period,
with (a) fringe contrasts and (b) shifts. 
The ratios of the Rabi frequencies are held constant, 
and the peak values of $\Omega_0/2\pi$ 
are shown in the horizontal axes as representatives of the Rabi
frequencies.  
$\alpha=0.55$ was used for the results
given in Fig.\ \ref{fig:robustness},
which gave relatively high contrasts for the case of $\Omega_2/2\pi\sim20$ kHz.
There is a region (above 10$\sim$30 kHz) over which 
the absolute values of the fringe contrasts are almost independent of the Rabi frequency,
which demonstrates the robustness.
The fringe shifts are expected to start from zero and increase
quadratically
as the Rabi frequency is
increased
due to the increase of AC Stark shifts.
The observed shifts shown in Fig.\ \ref{fig:robustness} (b) 
basically follow this expectation.

The results for the latter experiment are shown in Fig.\
\ref{fig:robustness}(c) and (d) with (c) fringe contrasts and (d) shifts.  
The fringe contrasts do not change appreciably
as the peak Rabi frequency is varied in the region above 20 kHz.  

We have performed numerical simulations trying to explain the loss of
fringe contrasts and the shifts.
Simulations are performed using a Liouville equation 
with decay terms representing
laser-frequency and magnetic-field flucutuations.
AC Stark shifts due to different Zeeman components and 
electric-dipole-allowed transitions 
($S_{1/2}$--$P_{1/2}$, $S_{1/2}$--$P_{3/2}$ and $D_{5/2}$--$P_{3/2}$)
are fully taken into account as
time-dependent variation of the detunings. 
The results are shown as curves in 
Fig.\ \ref{fig:robustness}, where
overall qualitative and partial quantitative agreements 
between experimental and numerical results
are obtained.


\section{Discussions}
\label{sec:discussions}

The possible factors for loss of fidelity in the previously described results
includes
laser frequency and magnetic field fluctuations.
The laser linewidth in the current setup is $\sim{}300$ Hz and the
magnetic field fluctuations corresponds to fluctuations of resonance 
frequencies of up to $\sim{}30$ Hz.
The effects of thermal distribution of motional quantum numbers
and intensity fluctuations are negligible, because the scheme used here
is robust against variations in Rabi frequencies.

We performed numerical simulations (similar to what is described in the
previous section) to evaluate the fidelity of 
some of the gates given in in \S \ref{sec:fidelity}. For X-rotation by angle
$\pi$ and Z-rotation by angle $\pi$, the dominant infidelity
factor was the effect of magnetic field fluctuations ($\sim$1 \%).  
Infidelity due
to diabaticity was estimated to be $\sim$0.2\%.  The contribution from
laser-frequency fluctuations in this case was below 0.1 \%. This can be
explained by the fact that the scheme is not sensitive to one-photon
detunings but only to two-photon detunings (as will be described later
in this section).  The contribution from laser-frequency fluctuations
may be larger when the encoding as in \S \ref{sec:rf} is used.
AC Stark shifts were compensated in this case by detuning one of the
beams as described before. 
The possible fidelity loss from not performing compensation
would amount to 20--30 \% in our simulation.

For the results of X- and  Z-rotations by angle $\pi/2$, 
the largest infidelity factor was diabaticity (2--5 \%),
and the effect of magnetic field fluctuations was 1--3 \%.
Although no AC-Stark-shifts compensation was performed in this case,
infidelity due to AC Stark shifts was as small as 0--1 \%,
and the effect of laser-frequency fluctuations was 0.2--0.3 \%.

%
%

The scheme demonstrated here is useful not only for atomic systems (neutral or
ionic), but also for other systems including solid-state systems.  The
scheme is effective for those systems where the coherence time is
relatively long, but inhomogeneities of the excitation field intensities cannot be avoided, or for those where slow intensity fluctuations of
excitation fields occur.  It can also be used effectively in assuring
equal operations in a large ensemble of particles.  The requirements
for the scheme are that a sufficient number of levels are available for
implementation, and that energy shifts due to excitation fields (such as
AC Stark shifts) can be avoided.

Elimination of AC Stark shifts is not a straightforward task for the
scheme presented here since a number of fields with time-dependent
amplitudes are used.  It can be confirmed from numerical simulations 
that the operations of the STIRAP gates are not sensitive to one-photon
detunings (i.e. detunings for
$\left|i\right\rangle$--$\left|u\right\rangle$ where $i=0,1$ and $2$)
but to every two-photon detuning, and hence all relative shifts between
two of $\left|0\right\rangle$, $\left|1\right\rangle$ and
$\left|2\right\rangle$ should be minimized. This might be relatively
complicated for {\it e.g.} $^{40}$Ca$^+$ since, in
addition to adjacent Zeeman components in $S_{1/2}$--$D_{5/2}$,
dipole-allowed transitions such as $S_{1/2}$--$P_{1/2}$,
$S_{1/2}$--$P_{3/2}$ and $D_{5/2}$--$P_{3/2}$ also give rise to
appreciable AC Stark shifts\cite{Haffner2003}.  

The scheme demonstrated here realizes
single-qubit operations that are in general noncommutable to each other
by using adiabatic manipulation of dark states.
These amount to implementations of noncommutable (non-Abelian) holonomies,
which can be a building block of HQC by Zanardi {\it et
al.}\cite{Zanardi1999} and leads to application in other fields
including neutral-atom systems \cite{Osterloh2005}.

\section*{ACKNOWLEDGEMENTS}
This work was supported by 
the Kakenhi "Quantum Cybernetics" project 
of the Ministry of Education, Culture, Sports, Science and Technology
(MEXT) in Japan and
the Japan Society for the Promotion of Science (JSPS) through its
Funding Program for World-Leading Innovative R\&D on Science and
Technology (FIRST Program).

%
%


\end{document}